\documentclass[conference]{IEEEtran}
\usepackage{cite}
\usepackage{graphicx}

\ifCLASSINFOpdf
  \usepackage[pdftex]{graphicx}
\else
\fi
\usepackage{url}

\begin{document}
%
\title{Million Atom Electronic Structure and\\Device Calculations on Peta-Scale Computers}

\author{\IEEEauthorblockN{%
 Sunhee Lee\IEEEauthorrefmark{1}, Hoon Ryu, Zhengping Jiang and Gerhard Klimeck 
}
\IEEEauthorblockA{\IEEEauthorrefmark{1} Network for Computational
Nanotechnology, Purdue University, West Lafayette, IN 47907, USA\\
e-mail: sunnyleekr@purdue.edu}}


%


\maketitle

\begin{abstract}
Semiconductor devices are scaled down to the level which constituent
materials are no longer considered continuous. To account for atomistic
randomness, surface effects and quantum mechanical effects, an atomistic
modeling approach needs to be pursued. The Nanoelectronic Modeling Tool
(NEMO 3-D) has satisfied the requirement by including emprical
$sp^{3}s^{*}$ and $sp^{3}d^{5}s^{*}$ tight binding models and
considering strain to successfully simulate various semiconductor
material systems. Computationally, however, NEMO 3-D needs significant
improvements to utilize increasing supply of processors. This paper
introduces the new modeling tool, OMEN 3-D, and discusses the major
computational improvements, the 3-D domain decomposition and the multi-level
parallelism. As a featured application, a full 3-D parallelized
Schr\"odinger-Poisson solver and its application to calculate the
bandstructure of $\delta$ doped phosphorus(P) layer in silicon is demonstrated. 
Impurity bands due to the donor ion potentials are computed. 
\end{abstract}


%
\IEEEpeerreviewmaketitle

\section{Introduction}
\emph{Need for Atomistic Modeling}: 
As semiconductor structures are scaled down to deca-nano sizes the
underlying material can no longer be considered continuous. The number
of atoms in the active device region becomes countable in the range of
50,000 to around 1 million and their local arrangement, becomes critical
in interfaces, alloys, and strained systems.  An atomistic modeling
approach needs to be used to capture such discreteness and quantum mechanical effects.  Most experimentally relevant structures are not infinitely periodic, but are finite in size and contain contacts; such geometries call for a local orbital basis, rather than a plane wave basis which implies infinite periodicity.   Furthermore we are primarily interested in stable semiconductor structures with well-established bonds which lessens or even eliminates the requirements to be able to compute the establishment of bonds with a full \emph{ab-initio} methodology.   

\emph{Multi-Million Atom Simulations}: 
NEMO 3-D \cite{GK00, GK01} uses empirical $sp^3s^*$ and $sp^3d^5s^*$ tight binding models that have been carefully calibrated to bulk materials in the III-V \cite{35param} and Si/Ge \cite{Si_Ge_param, SiGe_strain_param} material systems under various bulk strain and composition configurations.  This bulk parameterization is transferred to the nanoscale under the assumption of weak charge redistributions.   Weak piezo-electric effects in the InGaAs system can be captured through strain derived charge and electrostatic potential corrections\cite{GK02,Shaikh}. Transferability of the bulk parameters to nanometer devices was demonstrated by experimentally verified multi-million atom calculations for valley splitting in Si on SiGe\cite{Kharche}, single impurities in Si FinFETs, and InAs quantum dots in an InGaAs buffer matrix\cite{Usman}.  In these simulations none of the bulk parameters were modified and the nominal device dimensions were used to obtain quantitative agreement with experiment.  These simulations also showed that it was essential to include millions of atoms in the simulation domain and that simplified effective mass models have led to the wrong conclusions.

\emph{Computational Cost}: 
Multi-million atom calculations in NEMO 3-D come, however, at a typical computational price of 4-10 hours runtime on 20-64 cores on a standard cluster for a single evaluation of the eigenvalue spectrum.   Inclusion of this one pass electronic structure calculation into a self-consistent Poisson solution is possible, but drives the computation time up by another factor of 6-20.   This drives the computational requirement into the realm of days, rather than hours.   In sight of huge investments into Peta-Scale computing with availabilities of over 100,000 cores on a single supercomputer, efficient parallelism lays the goal for computational speed-up.  We have been able to demonstrate NEMO 3-D scaling to 8,196 processors\cite{NEMOscale}, however, such high level of scaling can only be achieved for unrealistically long essentially 1D structures, due to the 1D spatial parallel decomposition of NEMO 3-D.  NEMO 3-D therefore needs significant improvements in its parallelization schemes, data handling, post-processing, and code maintainability.

\emph{OMEN 3-D}: The major purpose of developing the new nanoelectronic
modeling tool, OMEN 3-D, emerged out from the need for expandability in growing
processor-rich environment. OMEN 3-D is equipped with more powerful
parallelization engine, 3-D domain decomposition scheme and general multi-level
parallelism. In addition, self-consistent charge calculations that need
additional
computational power are built in to deliver various kinds of
scientific simulations, from impurity physics to device applications.

This paper is organized as follows. In Sections II-A and B, the
parallelization schemes in OMEN 3-D and its benchmark
results are presented. In Section III-C, the multi-level parallelism is
briefly introduced. 
A Schr\"odinger-Poisson solver is
explained in Section III-A. Section III-B contains the example
of self-consistent bandstructure simulation on 2-D P
$\delta$-doped layer in silicon at 4K.

\section{Parallelization Scheme in OMEN 3-D}
The major feature in OMEN 3-D is its enhanced parallelization algorithm. 
NEMO 3-D uses a 1-D spatial decomposition scheme for parallelism. NEMO
3-D has been tested in many supercomputers and it is proven to show
close to perfect scalability, however, the maximum utilizable processors is strongly limited by its geometry in the 1D decomposition.

\begin{figure}
\centering
\includegraphics[width=0.70\columnwidth]{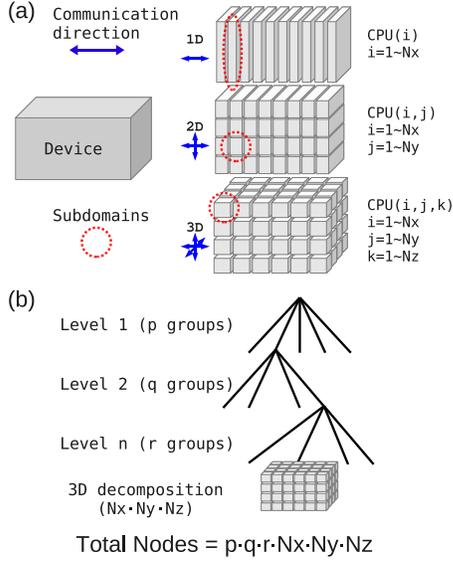}
\caption{(a) Schematic diagram of domain decomposition scheme. (b)
Multi-level parallelism in OMEN 3-D.}
\label{parallel}
\end{figure}

\subsection{3-D Spatial Domain Decomposition}
To reduce compute times by utilization of computers in excess of 10,000 cores, a new domain decomposition scheme is introduced in OMEN 3-D. 
In OMEN 3-D, a device of any shape can be spatially decomposed into
three dimensions and each subdomain is assigned to corresponding
processor. The maximum number of processors equals to the number of unit
cells in each direction.
Based on the spatial information, each processor only has the
list of information of the atoms in its subdomain and neighbor
atoms from adjacent subdomains; no global position infromation is held
locally, minimizing the memory consumption and making it possible to simulate
large devices (Fig.\ref{parallel}(a)).

The major drawback for 3D parallelization is the increase of the complexity of communication by $O(n^{N_{Dim}})$. The increased coupling among
the processors may cause significant performance degradation; 
there is a trade-off between reducing the computational burden and increasing communication overhead. 
However, recent benchmark results indicated the average time consumed
in the Message Passing Interface (MPI) communication is typically 5\% of the total simulation time.   
Moreover, from the NEMO 3-D cases, it was shown that the total simulation time was not bound by communication 
as long as the ratio of the number of surface atoms to the total number
of atoms in each subdomain is kept sufficiently small\cite{GK03}.

\subsection{Benchmark Results}

\begin{figure}
\centering
\includegraphics[width=0.80\columnwidth]{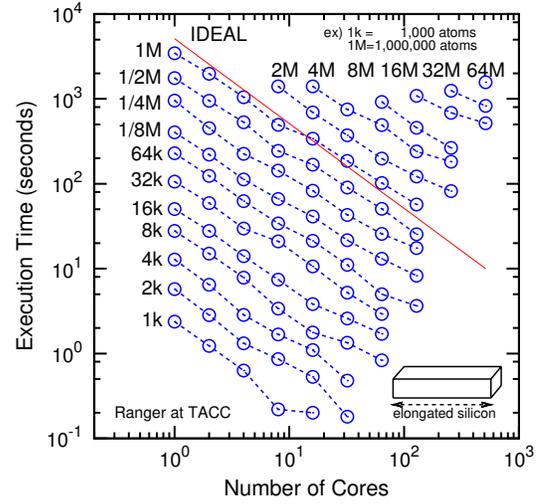}
\caption{Strong scaling benchmark results of a 1D Decomposition scheme
in OMEN 3-D. 500 Lanczos iterations are measured on elongated silicon
structures (subfigure). The number of atoms range from 1,000(`1k' in the figure) to
64 million (`64M')}
\label{fig_1d_decomp}
\end{figure}

\begin{figure}
\centering
\includegraphics[width=0.75\columnwidth]{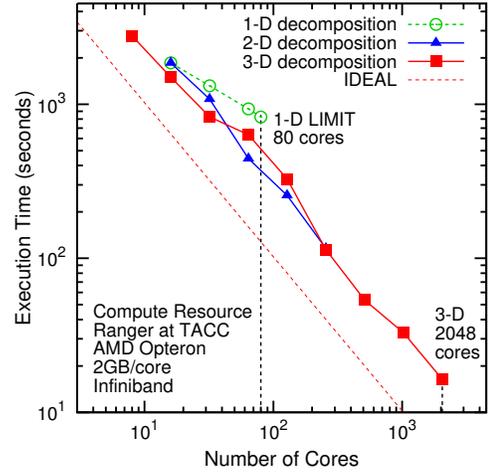}
\caption{Strong scaling comparison between 1/2/3D spatial decomposition. 
Performance of 500 Lanczos iterations is measured on a
$\mathrm{44\times44\times44(nm^{3})}$ silicon cube (4 million atoms). 
For the 2D case, the processors are assigned as ($\mathrm{c_{x},c_{y},c_{z}}$)=(16,$2^{i}$,1), $i=0,\cdots,4$.
And for 3D case, ($\mathrm{c_{x},c_{y},c_{z}}$)=($2^{i},2^{j},2^{k}$), $i,j,k=1,2,3$.
}
\label{fig_3d_decomp}
\end{figure}

\begin{figure}
\centering
\includegraphics[width=0.8\columnwidth]{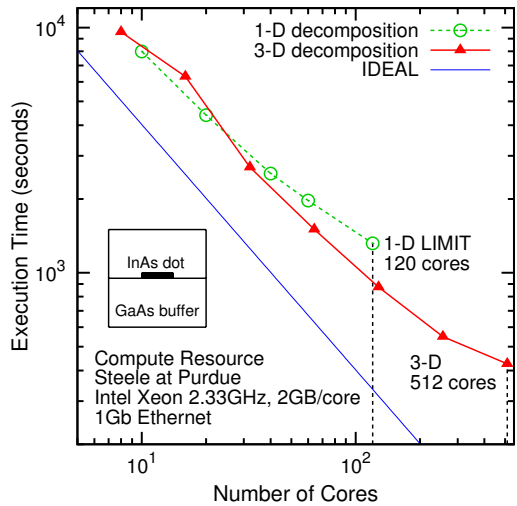}
\caption{Comparison of strain performance between 1D and 3D
decomposition. A cylindrical InAs QD of size 20nm(D)$\times$5nm(H) is
encapsulated in 68$\times$68$\times$68($\mathrm{nm^{3}}$) GaAs buffer.
This structure has 13 million atoms.
}
\label{fig_strain}
\end{figure}

The strong scaling plot of the 500 Lanczos iterations using the basic 1-D
parallelism for elongated systems of different number of atoms is presented in Fig.~\ref{fig_1d_decomp}. 
This plot indicates that with minimal load of communication, OMEN 3-D shows
reasonable scalability up to the structure that contains 32 million atoms with 512 processors in Ranger.  
However, with smaller number of atoms per subdomain, fluctuations in
performance are observed as we increase the number of processors; this
instability stems from the communication load being comparable to the computational operations.

The strong scaling plot in Fig.~\ref{fig_3d_decomp} examines the
performance of the 3-D decomposition scheme using 500 Lanczos iterations. 
The structure under test is a $\mathrm{44 \times 44 \times 44 (nm^{3})}$ silicon cube, which has 4 million atoms (80 unit cells in each direction). 
As with the previous strong scaling result, 1-D
decomposition scheme scales linearly; the number of processors can be
assigned is limited to 80. On the other hand, 2-D and 3-D
parallelization enables to assign more processors to the calculation,
resulting in a proportional time reduction. It is measured in this example that the performance is enhanced by 13.3 times by using 16 times more processors.
Therefore, by utilizing more computational resources, the 3-D
decomposition scheme opens the possibility of delivering simulation
results of realistic devices within significantly reduced time.

Typical NEMO 3-D simulations of InAs/GaAs Quantum Dot(QD) systems
\cite{Usman} not only involve electronic structure calculations but also
require minimization of the total strain energy in an atomistic Valence Force
Field (VFF) method \cite{GK01}. This strain calculation is
computationally significantly simpler than the subsequent electronic
structure calculation. It therefore does not in general scale as well with
increased parallelism. Here we test the VFF algorithm in 1D and 3D
decomposition in OMEN 3-D (Fig.~\ref{fig_strain}).
The sample structure is a cylindrical InAs QD of size
$\mathrm{20nm(D)\times5nm(H)}$ embedded in
$\mathrm{68\times68\times68(nm^{3})}$ GaAs buffer, 
which is comprised of 13 million atoms.
Again, 3-D decomposition scheme helps to scale down further to a factor of 3.5 by allocating 4 times more processors.

\subsection{Multi-Level Parallelism}
OMEN 3-D also has a programmable interface ready for multi-level
parallelism as depicted in Fig.~\ref{parallel} (b) to achieve extra performance enhancement. 
In contrast to the spatial domain decomposition, where the processors are coupled to each other by MPI communication, this hierarchical parallelism solves the task independently, with different parameters assigned for each group. 
K-space grouping, for example, can be useful when bandstructure or
charge calculations are needed. Additional bias groups can be added for
simulations that may involve external electrical or magnetic fields. Depending on the application, OMEN 3-D can provide multiple levels of additional parallelism to utilize more computational resources.
 
\section{Application}
\subsection{The Schr\"odinger-Poisson Solver}
One of the first applications of OMEN 3-D is the self-consistent charge and potential calculation module, known as the Sch\"odinger-Poisson solver, which was not present in NEMO 3-D.
There are three main components in the self-consistent loop:
\begin{enumerate}
    \item { \emph{Schr\"odinger Equation Solver}: Solves the eigenstates of the Schr\"odinger equation on finite k-points based on either $sp^{3}d^{5}s^{*}$ tight binding or effective mass Hamiltonian using iterative eigenvalue solver, such as, (block) Lanczos or PARPACK.}
    \item { \emph{Charge Calculation}: Based on the eigensolutions from
the Schr\"odinger equation, there are two different approaches to obtain
the charge profile. In the case of a given Fermi level, the charge profile can be calculated simply by filling up the states. On the other hand, if the Fermi level needs to be determined by external conditions, such as charge neutrality, both the charge and the Fermi level can be determined simultaneously.}
    \item { \emph{Poisson Solver}: Charge is fed into the Poisson
solver. The Poisson solver in OMEN 3-D also adopts 3-D parallelism and
uses a finite difference method with the Aztec linear solver. The
converged potential profile is iteratively obtained using
Newton-Raphson's method. The potential result is updated in the
Hamiltonian and the steps are repeated until the self-consistency is achieved.}
\end{enumerate}

The Schr\"odinger-Poisson solver has been applied to a couple of physical simulations. 
\begin{itemize}
    \item { \emph{Investigation of the Charge Distribution of a
Realistically Sized FinFET using the Top of the Barrier Model}
\cite{TOB00}\cite{TOB01}: The non-uniform current distribution in the
tri-gated devices of cross-section $\mathrm{65nm(H) \times 25nm(W)}$
versus the gate-voltage was successfully demonstrated.}
    \item { \emph{Bandstructure Calculation of 1D/2D Highly P
Doped silicon Structures} \cite{QC00}\cite{QC01} : Self-consistent bandstructure of closely positioned impurity atoms can be obtained. The self-consistent scheme applied to the impurity system is briefly introduced in the next section.}
\end{itemize}

\begin{figure}
\centering
\includegraphics[width=0.7\columnwidth]{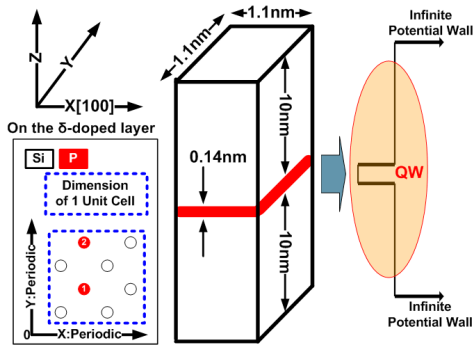}
\caption{The example structure of Si:P $\delta$ layer(red) embedded in 20nm
silicon buffer. It is periodic in 2D with planar doping of 1/4ML
($\mathrm{2.0 \times 10^{14} cm^{-2}}$).
}
\label{sim_device}
\end{figure}

\begin{figure}
\centering
\includegraphics[width=0.75\columnwidth]{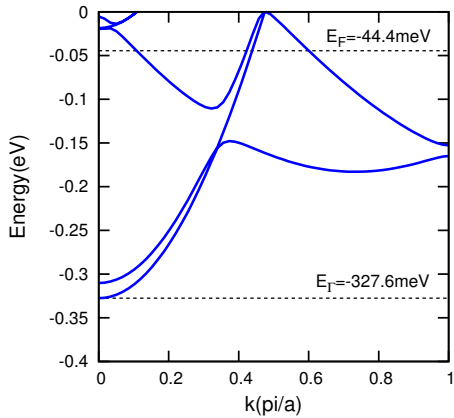}
\caption{The bandstructure result with respect to the silicon bulk
conduction band minima after self-consistency is achieved.
}
\label{T4K}
\end{figure}

\subsection{Example: Bandstructure of the Phosphorus $\delta$ layers in silicon}
Advances in fabrication process has enabled the creation of atomic-scale devices in
silicon. Using Scanning Tunneling Microscope (STM), experimentalists can
fabricate phosphorus $\delta$ layers and encapsulate them in silicon
\cite{QC02}.
This technology is significant in two fold; it is relevant in nanoelectronic device fabrication 
and it highlights the possibility of fabricating quantum computers.
Using the self-consistent method, the bandstructure of the 2-D periodic
$\delta$ layer structure with the doping density of 1/4 Mono Layer(ML)
(Fig.~\ref{sim_device}) is calculated at T=4K. The
bandstructure (Fig.~\ref{T4K}) indicates that due to the potential
induced by closely placed ionized donors, impurity bands are formed
below the silicon bulk conduction band. The band minima and the Fermi
level is located at 327.6meV and 44.4meV below the conduction band minima, respectively.  
For detailed simulation and analysis of the temperature dependence on
Si:P $\delta$ layer, refer to reference \cite{HR00}.

\section{Conclusion}
The new nanoelectronic simulator OMEN 3-D is developed to overcome the limitations of NEMO 3-D in a processor-rich environment. 
The new parallel algorithm introduced in OMEN 3-D shows better scalability
and is applicable to massive simulations and we expect to run further
tests on several thousands of processors. This work will allow us to perform NEMO 3-D like calculations in minutes rather than days. 
As an example, the Schr\"odinger-Poisson module and its application to
Si:P $\delta$ layer at 4K was introduced. Due to the
potential formed by impurity ions, set of impurity bands are observed
below the Si bulk conduction band. According to the simulation result, the Fermi level was located
44.4meV below the conduction band minima for 1/4ML Si:P layer. 

\section*{Acknowledgment}
NSF-funded nanoHUB.org and Ranger@TACC computational resources were used in this work.  
This work was supported by NSF, Purdue Research Foundation, and the Army Research Office.
Discussions with Mathieu Luisier, Benjamin Haley and Abhijeet Paul are gratefully acknowledged. 

\bibliographystyle{IEEEtran}
\bibliography{IEEEfull,IEEEexample}
%


\end{document}